\documentclass[epj,final]{svjour}
\usepackage{epsfig}

\makeatletter
\newbox\slashbox \setbox\slashbox=\hbox{\large$/$}
\def\pslash#1{\setbox\@tempboxa=\hbox{$#1$}
  \@tempdima=0.5\wd\slashbox \advance\@tempdima 0.5\wd\@tempboxa
  \copy\slashbox \kern-\@tempdima \box\@tempboxa}
\def\slash{\protect\pslash}
\def\openone{\leavevmode\hbox{\small1\kern-3.3pt\normalsize1}}
\makeatother

\begin{document}

\title{Statistical properties at the spectrum edge of the QCD Dirac
  operator} 
\author{Jian--Zhong Ma\inst{1}\and Thomas Guhr\inst{1}\and 
  Tilo Wettig\inst{2}}
\institute{Max--Planck--Institut f\"ur Kernphysik, Postfach 103980,
  D-69029 Heidelberg, Germany \and
  Institut f\"ur Theoretische Physik, Technische Universit\"at
  M\"unchen, D-85747 Garching, Germany}
\date{26 January 1998}

\abstract{%
  The statistical properties of the spectrum of the staggered Dirac
  operator in an SU(2) lattice gauge theory are analyzed both in the
  bulk of the spectrum and at the spectrum edge.  Two commonly used
  statistics, the number variance and the spectral rigidity, are
  investigated.  While the spectral fluctuations at the edge are
  suppressed to the same extent as in the bulk, the spectra are more
  rigid at the edge.  To study this effect, we introduce a microscopic
  unfolding procedure to separate the variation of the microscopic
  spectral density from the fluctuations.  For the unfolded data, the
  number variance shows oscillations of the same kind as before
  unfolding, and the average spectral rigidity becomes larger than the
  one in the bulk.  In addition, the short-range statistics at the
  origin is studied.  The lattice data are compared to predictions of
  chiral random-matrix theory, and agreement with the chiral Gaussian
  Symplectic Ensemble is found.}


\maketitle

\section{Introduction}
\label{sec1}

The spectrum of the Dirac operator is an important aspect of
nonperturbative QCD.  In Euclidean space, the Dirac operator reads
$i\slash D=i\slash\partial+g(\lambda^a/2)\slash A^a$, where $g$ is the
coupling constant, $\lambda^a$ are the generators of SU($N$)-color,
and $A_\mu^a$ are the gauge fields.  Whether in the continuum or on
the lattice, to obtain physical observables one has to perform an
average over the ensemble of gauge field configurations.  The
eigenvalues of $i\slash D$ fluctuate over this ensemble of gauge
fields.  In many areas of physics, e.g., in classically chaotic
systems, complex nuclei, and disordered mesoscopic systems, the study
of energy level statistics plays a prominent role.  Although the
global spectral fluctuations are strongly system dependent, the local
spectral correlations on the scale of the mean level spacing are
universal and only depend on the global symmetries of the system under
consideration.  These universal correlations can be obtained exactly
in random-matrix theory (RMT) \cite{Bohi84}.  In order to describe the
universal statistical properties of the eigenvalues of $i\slash D$,
one needs random-matrix models which take the chiral symmetry of the
Dirac operator into account \cite{Shur93}.  In a chiral random-matrix
model, the matrix of the Dirac operator (in a chiral basis in
Euclidean space) is replaced by a random matrix $W$ with appropriate
symmetries,
\begin{equation}
  \label{eq1.0}
  i\slash D \rightarrow \left[\matrix{0&W\cr W^\dagger&0}\right]\;.
\end{equation}
The average over gauge fields is replaced by an average over the
ensemble of random matrices, and the gluonic part of the weight
function, $\exp(-S_{\rm gl})$, is replaced by a simple Gaussian
distribution of the random matrix $W$.  Depending on the number of
colors and the representation of the fermions, the Dirac operator
falls into one of three universality classes which were classified in
Ref.~\cite{Verb94a} and correspond to the three chiral Gaussian
random-matrix ensembles.  These are the chiral Gaussian orthogonal
(chGOE), unitary (chGUE), and symplectic (chGSE) ensemble in which the
random matrix $W$ is either real, complex, or quaternion real,
respectively.  For a recent review of important results, we refer to
Ref.~\cite{Verb97}.

Various quantities, in addition to the spectral correlation functions,
have been employed to describe the statistical properties of the
spectrum of a given system.  One uses the number variance and the
spectral rigidity, to be defined below, to measure long-range
correlations, whereas the distribution of spacings between eigenvalues
is used to probe short-range correlations in the spectrum.  A
statistical analysis of the spectrum of the Dirac operator was first
performed using data obtained by Kalkreuter in an SU(2) lattice gauge
theory with both Wilson and Kogut-Susskind fermions \cite{Hala95}, and
it was found that the lattice data were described by RMT.  Recently,
high-statistics lattice data obtained by Berbenni-Bitsch and Meyer
were analyzed.  Again, perfect agreement with RMT was found
\cite{Wilk97}.  In the bulk of the spectrum, the local spectral
fluctuation properties follow the predictions of the conventional
Gaussian ensembles, which in this case are identical to those of the
chiral ensembles \cite{Naga91}.

The QCD Dirac operator has a special symmetry which has important
implications for the edge of the spectrum.  Since the Dirac operator
anti-commutes with $\gamma_5$, its eigenvalues appear in pairs
$\pm\lambda$ leading to level repulsion at $\lambda=0$.  Based on an
analysis of Leutwyler-Smilga sum rules \cite{Leut92}, Shuryak and
Verbaarschot \cite{Shur93} conjectured that the so-called microscopic
spectral density of the Dirac operator, defined by
\begin{equation}
  \label{eq1.1}
  \rho_{s}(z)=\lim_{V\to \infty}\frac{1}{V\Sigma}\;
  \rho\left(\frac{z}{V\Sigma}\right)\;, 
\end{equation}
should be a universal quantity as well.  Here,
$\rho(\lambda)=$\linebreak  
$\langle\sum_i\delta(\lambda-\lambda_i)\rangle_A$ is the overall
spectral density of the Dirac operator, $V$ is the space-time volume,
and $\Sigma$ is the absolute value of the chiral condensate,
respectively.  According to the Banks-Casher formula,
$\Sigma=\pi\rho(0)/V$ \cite{Bank80}, the spectral density at zero
virtuality is proportional to the chiral condensate, therefore the
spectrum edge is of great interest for the understanding of the
spontaneous breaking of chiral symmetry in QCD.  The definition of
Eq.~(\ref{eq1.1}) leads to a magnification of the region of small
eigenvalues by a factor of $V\Sigma$, thus the microscopic spectral
density also provides information about the approach to the
thermodynamic limit.

If $\rho_{s}$ is a universal quantity, it should also be calculable in
chiral random-matrix theory (chRMT).  This was done for all three
chiral ensembles in the chiral limit \cite{Verb93,Verb94c,Naga95}.
Several pieces 
of evidence have lent support to the universality conjecture for
$\rho_{s}$.  We mention here the reproduction of Leutwyler-Smilga sum
rules \cite{Verb93}, the microscopic spectral density in an instanton
liquid model \cite{Verb94b}, the valence quark mass dependence of the
chiral condensate \cite{Chan95,Verb96a}, theoretical universality
proofs with regard to the probability distribution of the random
matrix \cite{Brez96,Nish96,Slev93}, and finite temperature
calculations \cite{Jack96b}.  Very recently, the conjecture was
verified directly by lattice calculations.  In Ref.~\cite{Berb97a},
the microscopic spectral density, the distribution of the smallest
eigenvalue, and the microscopic two-point correlator were constructed
from lattice data calculated by Berbenni-Bitsch and Meyer for an SU(2)
gauge theory with staggered fermions in the quenched approximation and
compared with the random-matrix predictions.  The agreement was
excellent.
 
Another very interesting area of application of chiral random-matrix
theory is the construction of schematic models for the chiral phase
transition at finite temperature and/or chemical potential
\cite{Jack96a,Wett96,Step96a,Step96b,Nowa96,Jani97a,Hala97a,Hala97b}.
It should be 
emphasized that while chRMT yields exact results for the spectral
correlations in the bulk of the spectrum and at the spectrum edge, the
results of these schematic models typically are not universal.
Nevertheless, such models can shed light on a number of important
issues, such as the nature of the quenched limit at finite chemical
potential \cite{Step96b}.  Other aspects concern the relation with
Nambu--Jona-Lasinio models \cite{Jani97a} and the U$_{\rm A}$(1)
problem \cite{Jani97b}.  In this paper, however, we shall not discuss
such models but concentrate on the universal properties of the
spectrum of the Dirac operator.

It is the goal of the present work to perform the statistical spectral
analysis at the spectrum edge for the same lattice data as in
Ref.~\cite{Berb97a} and to compare to the statistical properties in
the bulk.  Our aim is two-fold: (i) to test, for higher order
correlations, the conjecture that the spectral statistics of the QCD
Dirac operator are universal and described by chRMT, and (ii) to gain
an intuitive understanding of the statistical properties of the Dirac
spectra at the edge (at zero virtuality).  The difference between the
usual situation (normally in the spectrum bulk) and the present one is
the hard edge at $\lambda=0$ which breaks the translational invariance
of the spectrum in this region.

In Sec.~\ref{sec2} we give, for the microscopic region, the
theoretical predictions for the one- and two-point spectral
correlation functions of the chGSE and the expressions for the number
variance and the average spectral rigidity applicable to the present
situation.  Section~\ref{sec3} is devoted to the analysis of lattice
data with regard to the long-range statistics, as well as the one- and
two-point correlation functions.  We then perform, in Sec.~\ref{sec4},
a microscopic unfolding in order to obtain a better understanding of
the behavior of the long-range correlations obtained in
Sec.~\ref{sec3}.  In Sec.~\ref{sec5} we study the spacing
distribution in a specific case.  The last section is a summary and
also discusses the effect of lattice parameters on our results.

\section{Theoretical predictions of the chGSE}
\label{sec2}

The appropriate ensemble corresponding to staggered fer\-mions in SU(2)
is the chGSE \cite{Verb94a}.  A central object in RMT is the joint
distribution function of the eigenvalues of the matrix in
Eq.~(\ref{eq1.0}) which is obtained after diagonalizing the matrix and
computing the Jacobian.  If $W$ has $N+\nu$ rows and $N$ columns, then
the matrix in (\ref{eq1.0}) has $N$ positive eigenvalues
$\lambda_1,\ldots,\lambda_N$, $N$ negative eigenvalues
$-\lambda_1,\ldots,-\lambda_N$, and $\nu$ zero modes (we assume
$\nu\ge 0$).  The joint distribution function then reads
\cite{Verb94a}
\begin{eqnarray}
  \label{eq2.1}
  P(\lambda_{1},\ldots,\lambda_{N})&\sim&
  \prod_{j=1}^{N}\lambda_{j}^{2N_f+4\nu+3}e^{-2N\Sigma^{2}\lambda_j^2}
  \prod_{k<l}^{N}(\lambda_k^2-\lambda_l^2)^4\;,\nonumber\\
  &&
\end{eqnarray}
where we have omitted a normalization factor which ensures that $\int
d\lambda_1\cdots d\lambda_N\rho=1$. Here, $N_f$ is the number of
massless flavors, and $\nu$ corresponds to the topological charge.
For the quenched lattice data we consider, $N_f=0$ and $\nu=0$.

We now need expressions for the scaled one- and two-point functions at
the spectrum edge. They can be derived from results computed by Nagao
and Forrester \cite{Naga95} for the Laguerre ensemble with the joint
distribution function
\begin{equation}
  \label{eq2.2}
  P(x_{1},\dots,x_{N})\sim\prod_{j=1}^{N}x_{j}^{4a}e^{-4x_{j}}
  \prod_{k<l}^{N}(x_{k}-x_{l})^{4} 
\end{equation}
by a simple transformation of variables, $4x=2N\Sigma^2\lambda^2$.
Evidently, $4a=N_f+2\nu+1$. We use the same letter $P$ to
denote the joint distribution functions (\ref{eq2.1}) and (\ref{eq2.2}).
For convenience, we introduce a new parameter
\begin{equation}
  \mu=N_f+2\nu
\end{equation}
which will characterize the dependence of the various quantities we
consider on $N_f$ and $\nu$.  For the present lattice data we have 
$\mu=0$.  The $k$-point function follows by integrating the joint
distribution function over all but $k$ variables,
\begin{eqnarray}
  \label{eq2.3}
  R_k(\lambda_1,\ldots,\lambda_k)&=& \nonumber\\
  &&  \hspace*{-65pt}\frac{N!}{(N-k)!}
  \int\cdots\int d\lambda_{k+1}\cdots d\lambda_{N}
  P(\lambda_{1},\ldots,\lambda_{N}) \;.
\end{eqnarray}
A note on notation: We denote the usual $k$-point functions by $R_k$
and the corresponding microscopic limits by $\rho_k$.  The microscopic
limit of the $k$-point function is obtained by rescaling all arguments
by a factor of $2N\Sigma$ in analogy with Eq.~(\ref{eq1.1}).  Setting
the lattice constant to unity, we can identify the volume $V$ with
the number of eigenvalues $2N+\nu\approx 2N$ in the large-$N$ limit.
Including the proper normalization and using the result of
Ref.~\cite{Naga95}, we obtain for the microscopic one-point function
\begin{eqnarray}
  \label{eq2.4}
  \rho_s(z)&=&\lim_{N\to\infty}\frac{1}{2N\Sigma}
  R_{1}\left(\frac{z}{2N\Sigma}\right) \nonumber\\ 
  &=&2z^2\int_{0}^{1}duu^2\int_{0}^{1}dv[J_\mu(2uvz)J_{\mu+1}(2uz)
  \nonumber\\ 
  &&\phantom{2z^2\int_{0}^{1}duu^2\int_{0}^{1}}
  -vJ_{\mu}(2uz)J_{\mu+1}(2uvz)] \;,
\end{eqnarray}
where $J$ denotes the Bessel function.  Asymptotically, $\rho_{s}(z)$
$\to 1/\pi$ as $z\to \infty$.  To compare with the bulk properties, it
is convenient to rescale the argument by a factor of $\pi$ so that it
is measured exactly in terms of the local mean level spacing.  (This
factor of $\pi$ comes in through the Banks-Casher relation,
$\Sigma=\pi\rho(0)/V$.)  The result in Eq.~(\ref{eq2.4}) can be
simplified further and expressed in terms of a single integral
\cite{Berb97b}.  We obtain after some algebra
\begin{eqnarray}
  \label{eq2.5}
  \rho_1(z)&=&\pi\rho_{s}(\pi z)\nonumber\\
  &=&\pi^2 z[J_{\mu}^2(2\pi z)-J_{\mu+1}(2\pi z)J_{\mu-1}(2\pi z)]
  \nonumber\\
  &&-\frac{\pi}{2}J_{\mu}(2\pi z)\int_0^{2\pi z}dtJ_{\mu}(t) \;.
\end{eqnarray}
The two-point spectral correlation function is given by
\begin{equation}
  \label{eq2.6}
  R_{2}(\lambda_1,\lambda_2)=R_{1}(\lambda_{1})R_{1}(\lambda_{2})
  -T_{2}(\lambda_{1},\lambda_{2})\;,
\end{equation}
where $T_{2}(\lambda_{1},\lambda_{2})$ is the two-point cluster
function which contains the non-trivial correlations.  We write the
microscopic limit of Eq.~(\ref{eq2.6}) as
$\rho_{2}(z_1,z_2)=\rho_{1}(z_{1})\rho_{1}(z_{2})-\tau_{2}(z_{1},z_{2})$.
We are mainly interested in the microscopic limit of $T_2$, i.e., in
$\tau_2$.  Making use of the results of Ref.~\cite{Naga95}, we obtain
\begin{eqnarray}
  \label{eq2.7}
  \tau_2(z_1,z_2)&=&\lim_{N\to\infty}\left(\frac{\pi}{2N\Sigma}\right)^2
  T_2\left(\frac{\pi z_1}{2N\Sigma},\frac{\pi z_2}{2N\Sigma}\right)
  \nonumber\\
  &=&(2\pi^3z_1z_2)^2[S(\pi z_1,\pi z_2)S(\pi z_2,\pi z_1)\nonumber\\
  &&\phantom{(2\pi^3z_1z_2)}+I(\pi z_1,\pi z_2)D(\pi z_2,\pi z_1)]
\end{eqnarray}
with
\begin{eqnarray}
  S(x,y)&=&\int_{0}^{1}duu^2\int_{0}^{1}dv[J_{\mu}(2uvx)J_{\mu+1}(2uy)
  \nonumber\\ 
  \label{eq2.8a}
  &&\phantom{\int_{0}^{1}duu^2\int_{0}^{1}d}
  -vJ_{\mu}(2ux)J_{\mu+1}(2uvy)]\;,\\
  I(x,y)&=&\int_{0}^{1}duu\int_{0}^{1}dv[J_{\mu}(2uvx)J_{\mu}(2uy)
  \nonumber\\ 
  \label{eq2.8b}
  &&\phantom{\int_{0}^{1}duu\int_{0}^{1}dv[}
  -J_{\mu}(2ux)J_{\mu}(2uvy)]\;,\\
  D(x,y)&=&-\int_{0}^{1}duu^3\int_{0}^{1}dvv[J_{\mu+1}(2uvx)
  J_{\mu+1}(2uy) \nonumber\\ 
  \label{eq2.8c}
  &&\phantom{\int_{0}^{1}duu^3\int_{0}^1}
  -J_{\mu+1}(2ux)J_{\mu+1}(2uvy)]\;.
\end{eqnarray}
Again, this result can be simplified further and expressed in terms of
two single integrals \cite{Berb97b}.  We obtain
\begin{eqnarray}
  \label{eq2.9}
  \tau_2(z_1,z_2)&=&f(\pi z_1,\pi z_2)\partial_{z_1}\partial_{z_2}
  f(\pi z_1,\pi z_2)\nonumber\\
  &&+\partial_{z_1}f(\pi z_1,\pi z_2)\partial_{z_2}f(\pi z_2,\pi z_1)
\end{eqnarray}
with
\begin{equation}
  \label{eq2.10}
  f(x,y)=\frac{y}{2}\int_0^{2x}dtC(t,2y)-\frac{x}{2}\int_0^{2y}dtC(t,2x)
\end{equation}
and
\begin{equation}
  \label{eq2.11}
  C(x,y)=\frac{xJ_{\mu+1}(x)J_{\mu}(y)-yJ_{\mu}(x)J_{\mu+1}(y)}{x^2-y^2}\;.
\end{equation}
The derivatives of $f$ can be expressed as
\begin{equation}
  \label{eq2.12}
  \partial_xf(x,y)=2yC(2x,2y)-\frac{1}{2}J_{\mu}(2x)\int_0^{2y}dtJ_{\mu}(t)
\end{equation}
and
\begin{eqnarray}
  \label{eq2.13}
  \partial_x\partial_yf(x,y)&=&\frac{1}{x^2-y^2}\hspace*{50mm}
  \nonumber\\
  && \hspace*{-21mm}\times \Bigl\{
  (x^2+y^2)\bigl[2C(2x,2y)-J_\mu(2x)J_\mu(2y)\bigr]\nonumber\\
  && \hspace*{-16mm}+xy\bigl[J_{\mu+1}(2x)J_{\mu-1}(2y)
  +J_{\mu-1}(2x)J_{\mu+1}(2y)\bigr]\Bigr\}\,.
\end{eqnarray}

We wish to study the number statistics at the spectrum edge.  In the
study of spectral statistics, one usually has to unfold the empirical
spectrum in order to separate the global variations from the local
fluctuations, since the former are not universal and beyond the
predictions of RMT.  In the present case, the global spectral density
near the edge is constant to a good approximation, therefore no
unfolding is necessary in this region.  We simply rescale the energies
to introduce a dimensionless variable $S$ as in Eqs.~(\ref{eq2.5}) and
(\ref{eq2.7}).  Consider a small region $[0,L]$.  The average number
of eigenvalues within this region is
\begin{equation}
  \label{eq2.14}
  \bar{N}(0,L)=\int_{0}^{L}d\lambda R_{1}(\lambda) \;.
\end{equation}
Using the scale we set above, i.e., measuring the length of the
interval in terms of the local mean level spacing, we define
\begin{equation}
  \label{eq2.15}
  S=\frac{2N\Sigma}{\pi}L  
\end{equation}
and have
\begin{equation}
  \label{eq2.16}
  \bar{N}(0,S)=\int_{0}^{S}dz \rho_{1}(z)
\end{equation}
in the thermodynamic limit $N\rightarrow\infty$ with $S$ finite.  The
number variance is defined as $\Sigma^2(I)=\langle (N(I)-{\bar
  N}(I))^2\rangle$, where $I$ is a given interval and $N(I)$ is the
number of eigenvalues therein.  In the interval $[0,S]$, the number
variance can be expressed as
\begin{equation}
  \label{eq2.17}
  \Sigma^{2}(0,S)=\int_{0}^{S}dz
  \rho_{1}(z)-\int_0^S\int_0^Sdz_{1}dz_{2}\tau_2(z_{1},z_{2})\;.
\end{equation}
Note that $\Sigma^2$ should not be confused with $\Sigma$, the
absolute value of the chiral condensate.

Apart from the number variance, the spectral rigidity $\Delta_3$
introduced by Dyson and Mehta has played a major role in the study of
spectral statistics. It is defined as the mean-square deviation of the
cumulative level density $N(0,L)$ in an interval $[0,L]$ from
the best-fitting straight line,
\begin{eqnarray}
  \label{eq2.18}
  \Delta_{3}(0,L)&\equiv&\min_{A,B}\frac{1}{L} \int_{0}^{L}dx
  [N(0,x)-Ax-B]^{2}\nonumber\\
  &=&\langle N^{2}(0,x)\rangle-\langle N(0,x)
  \rangle^{2}\nonumber\\
  &&\hspace*{20pt}-\frac{12}{L^{2}}\langle (x-L/2)N(0,x)\rangle^2
\end{eqnarray}
with $\langle\cdots\rangle=\frac{1}{L}\int_{0}^{L}dx\cdots$ .  In RMT,
the averaged spectral rigidity can be expressed in terms of $R_1$ and
$R_2$.  Again, we are interested in a microscopic region at the
spectrum edge.  Using the scale defined in (\ref{eq2.15}), one has for
the chGSE in the microscopic region
\begin{eqnarray}
  \label{eq2.19}
  \bar{\Delta}_{3}(0,S)&=&\frac{1}{S^{4}}\int_{0}^{S}dz\rho_{1}(z)
  \left(S^{3}z-4S^2z^2+6Sz^{3}-3z^{4}\right)\nonumber\\
  &&\hspace*{-30pt}-\frac{2}{S}\int_{0}^{S}dz_1z_1\int_{0}^{z_1}dz_2
  \rho_{2}(z_1,z_2)\nonumber\\
  &&\hspace*{-30pt}+\frac{1}{S^{4}}\int_{0}^{S}\int_{0}^{S}dz_1dz_2
  \rho_2(z_1,z_2)[S^3(z_1\!+\!z_2)-4S^2z_1z_2\nonumber\\
  &&\hspace*{55pt}+3Sz_1z_2(z_1\!+\!z_2)-3z_1^2z_2^2]\;. 
\end{eqnarray}

For the number variance and the averaged spectral rigidity in an
arbitrary interval $[S_{0},S_{0}+S]$ we have similar expressions.
These theoretical predictions from the chGSE will be compared with
lattice data in Sec.~\ref{sec3}.  We will also compare the statistical
properties at the edge with those in the bulk, which should be
described by the GSE as mentioned in the introduction.  Explicit
expressions can be found in Ref.~\cite{Meht91}.

\section{Spectral correlators and long-range statistics}
\label{sec3}
      
We now analyze lattice data calculated by Berbenni-Bitsch and Meyer.
They computed complete spectra of the staggered Dirac matrix for an
SU(2) gauge theory using various values of $\beta$ and a number of
different lattice sizes.  We will focus on the data from an $8^{4}$
lattice with $\beta=2.0$.  Here, 3896 independent configurations were
obtained.  Because of the spectral ergodicity property of RMT, one can
construct the spectral correlations in the bulk with much fewer
configurations since the ensemble average can be replaced by a
spectral average \cite{Hala95}.  However, if one is interested in the
spectrum edge one has to perform an ensemble average, therefore a
large number of independent configurations is needed.

For the present spectra, we have computed the average global spectral
density and found that it can indeed be separated from the local
fluctuations by unfolding.  We also found that within the small
interval at the spectrum edge we are interested in, there is no
visible variation in the spectral density.  Therefore, no unfolding is
needed in the microscopic region.  However, we will perform a
``microscopic unfolding'' later, see Sec.~\ref{sec4}.  In order to
compare the data with the predictions of chRMT, only a simple
rescaling by the local mean level spacing is done in small intervals
starting at the spectrum edge.  A typical interval includes
approximately 20 eigenvalues.

\begin{figure}
  \centerline{\epsfig{figure=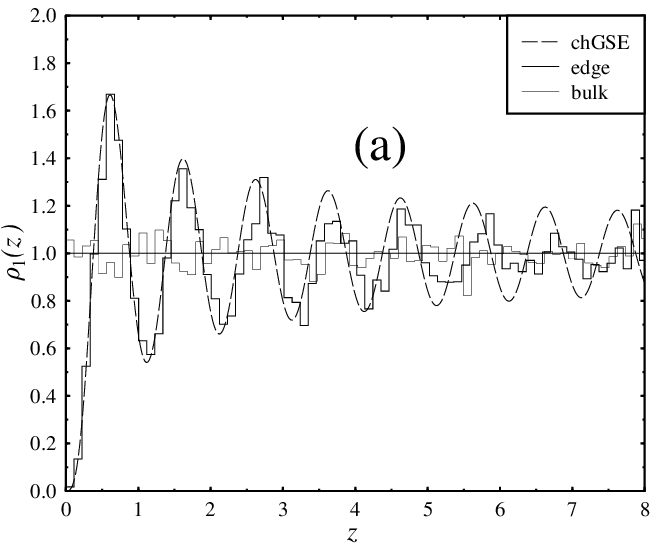,width=7.5cm}}
  \vspace*{5mm}
  \centerline{\epsfig{figure=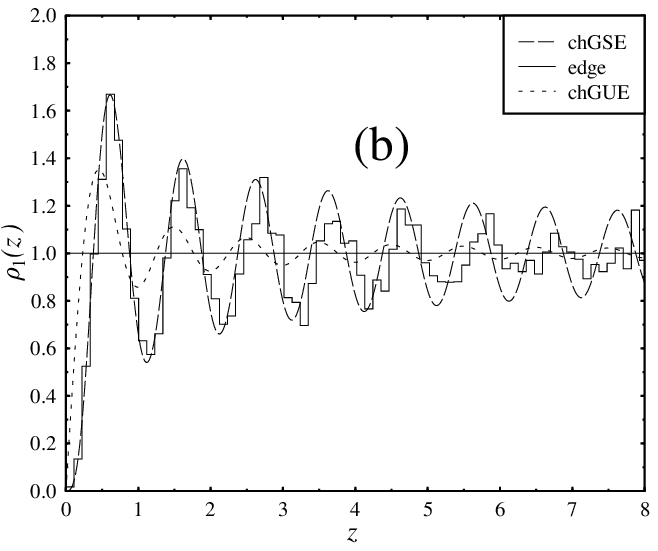,width=7.5cm}}
  \vspace*{3mm}
  \caption{The spectral density at the edge on the scale of the mean
    level spacing calculated from the lattice data and from the chGSE
    prediction is compared (a) with that in the bulk and (b) with the
    chGUE prediction.}
  \label{fig1}
\end{figure}

Figure~\ref{fig1} shows the results for the microscopic spectral
density.  We found that the agreement with the chGSE is very good for
$z<2$, quite impressive in the region $2<z<5$, and getting worse as
$z$ increases.  It is interesting to discuss what determines the
domain of validity of the random-matrix result.  The microscopic
spectral density at the edge is essentially
\begin{equation}
  \label{eq3.1}
  \rho_{1}(z)=\bar{D}(0)R_{1}(z\bar{D}(0))\;,
\end{equation}
where $\bar{D}(0)$ is the local mean level spacing at the edge.  In
order to obtain agreement between the lattice data and the
thermodynamic limit of the microscopic spectral density given by
Eq.~(\ref{eq2.5}), $z$ has to be small so that $ z\bar{D}(0)\ll 1$.
This is because by taking the thermodynamic limit in
Eq.~(\ref{eq2.4}), we treat the argument $\lambda$ of $R_1(\lambda)$
as an infinitesimal quantity.  Thus, the random-matrix result is valid
over a larger range if $\bar{D}(0)$ is small.  We have
$\bar{D}(0)=1/V\Sigma$, and the actual value of $\bar{D}(0)$ is
determined by several factors.  It decreases with increasing lattice
volume and increasing chiral condensate (everything is in lattice
units).  The condensate (at fixed lattice volume) depends on the
coupling constant and on temperature, and it decreases with both
$\beta$ and $T$.  Thus, the domain of validity of the random-matrix
result is smaller for smaller lattice volume, larger $\beta$, and
larger temperature.  To quantify these statements would require a
systematic study of lattice data at several values of $V$, $\beta$,
and $T$ which is beyond the scope of this paper.  For the present
lattice, we found $\bar{D}(0)\approx 0.0059$.  Therefore, agreement
with Eq.~(\ref{eq2.5}) is restricted to $z\ll 170$.  In practice, we
observe agreement only for values of $z$ much lower than this upper
bound. As we stressed earlier, the volume $V$ has to be identified
with $2N$, i.e., with twice the dimension of the matrices in the
random-matrix model. Of course, one can also construct $\rho_1$ for
purely random matrices of finite dimension.  In this case, the
agreement with Eq.~(\ref{eq2.5}) is quite good already for small
dimension.  In the case of the lattice data, however, the agreement
with Eq.~(\ref{eq2.5}) is worse for finite $V$ because the lattice
Dirac operator has additional non-random components.  We thus
attribute the disagreement between the lattice data and the prediction
of chRMT for $z>5$ to both the validity of the thermodynamic limit and
the non-random components of the lattice Dirac operator.

{}From Fig.~\ref{fig1}, we see that the microscopic spectral density has
an oscillatory pattern with peaks distributed almost periodically with
period $\approx 1$.  The position of the $i$-th peak is the most
probable value of the $i$-th eigenvalue.  The distribution of the
eigenvalues at the edge looks somewhat like a picket fence as does the
distribution of energy levels of a harmonic oscillator.  This is a
consequence of the strong repulsion of the eigenvalues with the fixed
point $z=0$.  The height of a peak decays as $z$ increases and
asymptotically tends to $1$. In Fig.~\ref{fig1}(a), the spectral
density in the bulk (around the midpoint of the spectra) is also
plotted for comparison.  There are only slight fluctuations around the
average value $\rho_{\rm bulk}=1$.  We also plot the chGUE result for
the microscopic spectral density \cite{Verb93},
\begin{equation}
  \label{eq3.2}
  \rho_1(z)=\frac{\pi^2}{2} z [J_0^2(\pi z)+J_1^2(\pi z)]
\end{equation}
for $N_f=\nu=0$, in Fig.~\ref{fig1}(b).  It can be seen that the peaks
of the chGUE are less pronounced and decay faster than those of the
chGSE due to its weaker repulsion (Dyson index $\beta=2$ vs
$\beta=4$).

\begin{figure}
  \centerline{\epsfig{figure=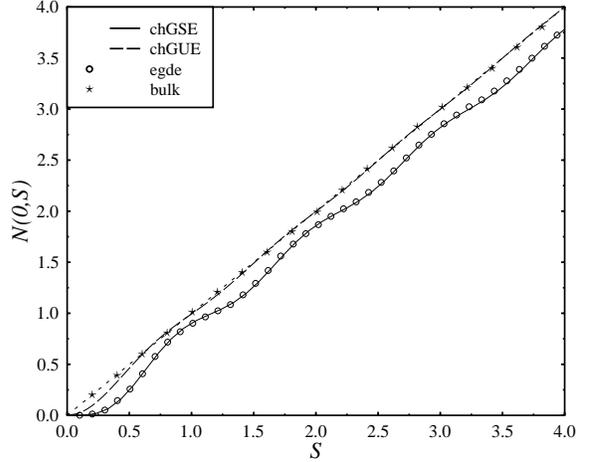,width=75mm}}
  \vspace*{3mm}
  \caption{The averaged staircase function at the spectrum edge and in
    the bulk. The full and the long-dashed line represent the chGSE
    and the chGUE, respectively. The short-dashed line is $\bar N=S$,
    representing the averaged staircase in the spectrum bulk.
    The symbols are the data.}
  \label{fig2}
\end{figure}

In Fig.~\ref{fig2}, we plot the averaged number of eigenvalues in the
interval $[0,S]$, i.e., the averaged staircase, as a function of $S$
at the edge as well as in the bulk.  Again, the agreement with the
chGSE, Eq.~(\ref{eq2.16}), is good.  The short-dashed line represents
the homogeneous distribution $\bar N=S$ in the bulk.  (Note that for
the lattice data we consider, the number of eigenvalues is large
enough so that for the small range of $S$ shown in Fig.~\ref{fig2},
unfolding in the bulk is not actually necessary.)  The deviation of
the chGSE from the straight line $\bar N=S$ at the edge is much larger
than that of the chGUE as in the case of the microscopic spectral
density.

\begin{figure}
  \centerline{\epsfig{figure=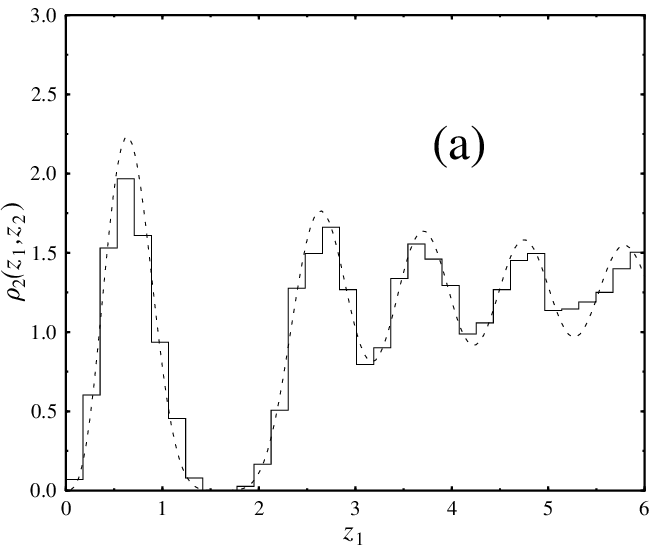,width=6.2cm}}
  \vspace*{3.4mm}
  \centerline{\epsfig{figure=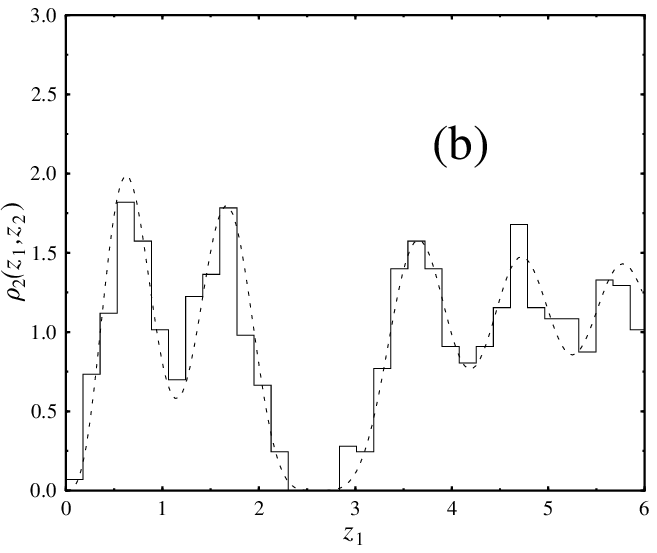,width=6.2cm}}
  \vspace*{3.4mm}
  \centerline{\epsfig{figure=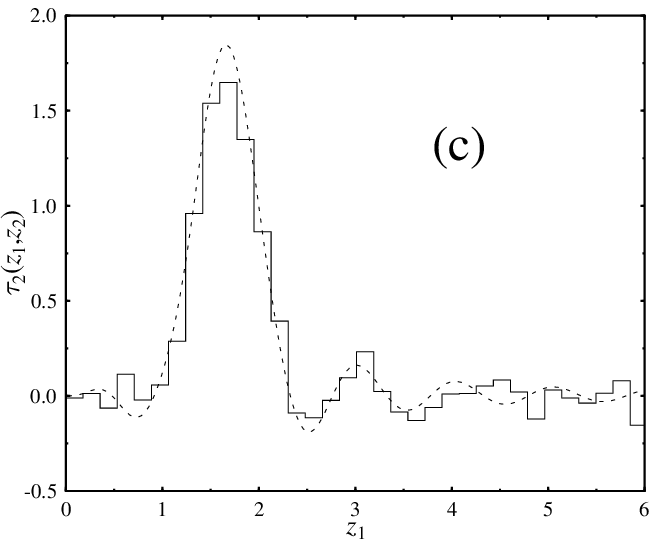,width=6.2cm}}
  \vspace*{3.4mm}
  \centerline{\epsfig{figure=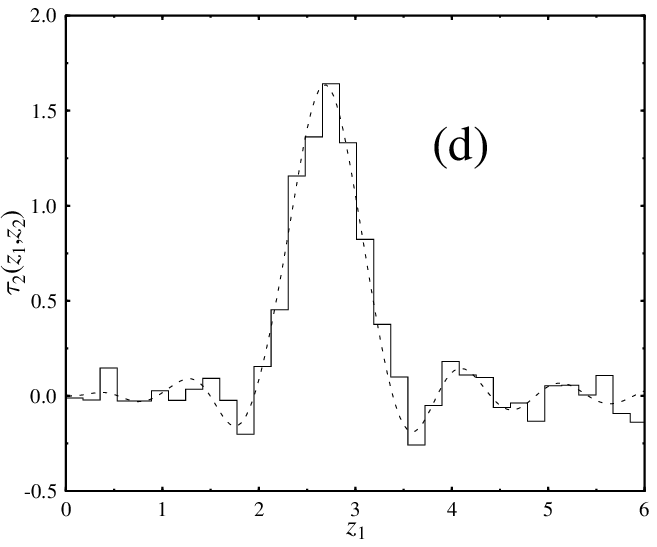,width=6.2cm}}
  \vspace*{0.5mm}
  \caption{The microscopic two-point correlation function
    $\rho_2(z_1,z_2)$ and the corresponding microscopic
    two-point cluster function $\tau_2(z_1,z_2)$ as functions of
    $z_1$ for fixed $z_2$.  In (a) and (c), $z_2=1.6$, in (b) and
    (d), $z_2=2.6$.  The histograms represent lattice data, the
    dashed lines the chGSE predictions.}
\label{fig3}
\end{figure}

We now turn to the two-point correlator. In Fig.~\ref{fig3} we plot
the results for $\rho_{2}(z_{1},z_{2})$ and $\tau_{2}(z_{1},z_{2})$ as a
function of $z_{1}$ for two fixed values of $z_{2}$, $z_{2}=1.6$ and
$z_2=2.6$.  One can see that the histogram for $\rho_{2}(z_{1},z_{2})$
agrees well with the chGSE prediction, although the agreement is not
as good as in Fig.~\ref{fig1} for the one-point function.  For
$\tau_{2}(z_{1},z_{2})$, the statistics are worse than for
$\rho_{2}(z_{1},z_{2})$, except for a small region around
$z_{1}=z_{2}$.  This is because $\rho_{2}(z_{1},z_{2})$ also includes
the one-point functions which are dominant and have better statistics.
{}From the definitions of $\rho_{2}(z_{1},z_{2})$ and
$\tau_{2}(z_{1},z_{2})$ we can see that one needs a much larger number of
spectra to obtain better statistics.  For a given value of $z_2$, only
those configurations in which there is an eigenvalue in the bin around
$z_2$ contribute to the two-point function.  Even for a value of
$z_{2}$ chosen at a peak of $\rho_{1}$ (as in Fig.~\ref{fig3}), no
more than $1/3$ of the configurations are actually involved in the
ensemble average in the construction of $\rho_{2}(z_{1},z_{2})$ and
$\tau_{2}(z_{1},z_{2})$ from the data.  This is the reason why we only
computed $\rho_{2}(z_{1},z_{2})$ and $\tau_{2}(z_{1},z_{2})$ for these
specific values of $z_{2}$.  For other values, one would have to
choose a larger bin size.

\begin{figure}
  \centerline{\epsfig{figure=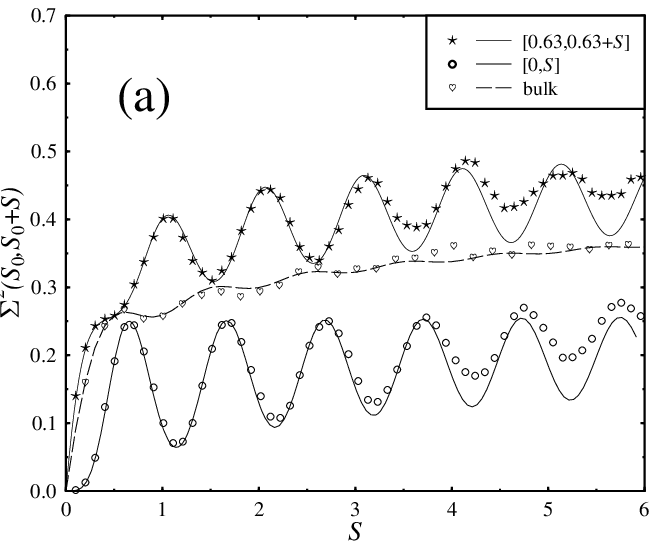,width=6.1cm}}
  \vspace*{3.4mm}
  \centerline{\epsfig{figure=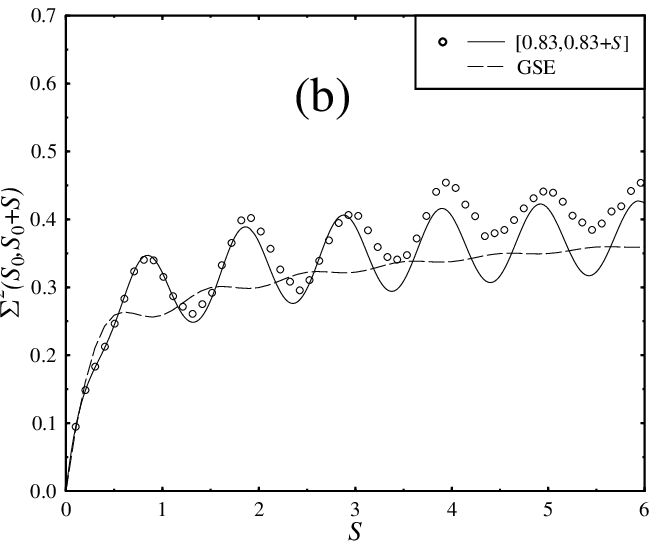,width=6.1cm}}
  \vspace*{3.4mm}
  \centerline{\epsfig{figure=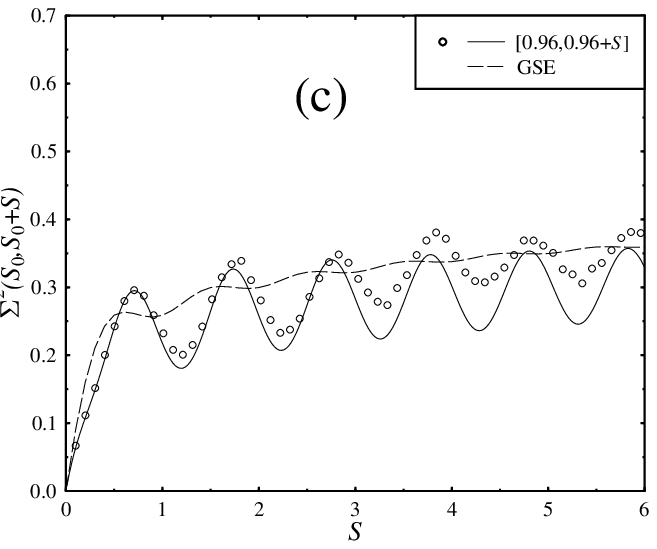,width=6.1cm}}
  \vspace*{3.4mm}
  \centerline{\epsfig{figure=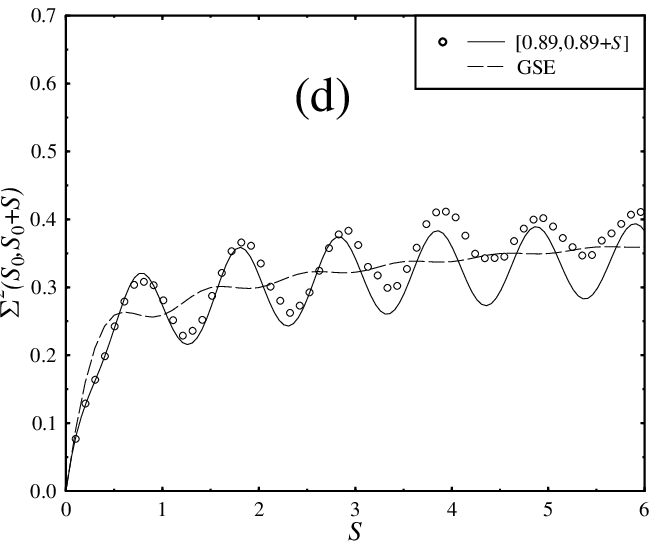,width=6.1cm}}
  \vspace*{0.5mm}
  \caption{The number variance in an interval $[S_0,S_0+S]$ at the edge 
    with (a) $S_0=0$ and 0.63, (b) $S_0=0.83$, (c) $S_0=0.96$, and (d)
    $S_0=0.89$ as calculated from the lattice data is compared with
    the corresponding predictions of the chGSE (solid lines) and GSE
    (dashed lines).  For comparison, the number variance in the bulk
    is computed in (a).}
  \label{fig4}
\end{figure}

Figure~\ref{fig4} shows our results for the number variance
$\Sigma^{2}(S_{0},S_{0}+S)$ at the edge for different values of
$S_{0}$ compared with the corresponding chGSE predictions. We also
calculated the number variance in the bulk for comparison in
Fig.~\ref{fig4}(a).  No spectral average was performed in the bulk.
{}From this figure, one observes the following points: (i) At the edge,
significant systematic deviations from the theoretical prediction
occur when $S_{0}+S>4$.  The data approach the asymptotic behavior
faster than the theoretical curve.  Again, this means that the
agreement with the thermodynamic limit is restricted to the small
region $[0,4\sim 5]$, cf.\ the discussion after Eq.~(\ref{eq3.1}).
(ii) For $S\ll 1$, the number variance at the edge increases very
slowly, in contrast to the linear relation $\Sigma^{2}=S$ in the bulk.
This is simply a manifestation of the suppression of the microscopic
spectral density in this region.  In fact, from Eq.~(\ref{eq2.17}) it
can be seen that $\Sigma^{2}$ is dominated by the first term, i.e.,
by the average staircase function, when $S\ll 1$.  The two-point
correlations manifest themselves in $\Sigma^{2}(S_{0},S_{0}+S)$ only
if $S_0+S\ge 1$.  (iii) The overall value of
$\Sigma^{2}(S_{0},S_{0}+S)$ strongly depends on the value of $S_{0}$.
The larger the value of $\rho_{1}(S_{0})$, the larger
$\Sigma^{2}(S_{0},S_{0}+S)$.  Two extreme cases are shown in
Fig.~\ref{fig4}(a) where the entire curves of
$\Sigma^{2}(S_{0},S_{0}+S)$ are higher resp. lower than the GSE curve.
The values $S_{0}=0.63$ and $S_{0}=0$ correspond to the first maximum
and minimum of $\rho_{1}$, respectively.  Figures~\ref{fig4}(b), (c),
and (d) show the results for $S_{0}=0.83$, 0.96, and 0.89 with
$\rho_{1}(S_{0})<1, >1, $ and $=1$, respectively.  (iv) For fixed
$S_{0}$, the curves show strong oscillations with peaks appearing when
$\rho_{1}(S_{0}+S)$ reaches its maxima.  The values of $\Sigma^{2}$
for arbitrary intervals $[S_{1},S_{2}]$ at the edge are distributed
around the GSE curve.  It is known \cite{Meht63} that the number
variance reflects the ``compressibility'' of the eigenvalue ``gas''.
Therefore, roughly speaking, the compressibility at the edge is on
average the same as in the bulk.  The case $S_{0}=0$ demands special
attention.  The very strong suppression shown in Fig.~\ref{fig4}(a) is
special because the interval starts at the origin which is a fixed
point for all spectra so that fluctuations of the eigenvalue number
from the left hand side are prohibited.  It is always harder to
compress this one-dimensional gas on one side than on two sides.

At first glance, one might attribute these features of $\Sigma^{2}$ to
the inhomogeneity of the microscopic spectral density.  To clarify
this point, we will in the next section introduce a microscopic
unfolding procedure to remove the oscillations in the microscopic
spectral density, and then investigate the effect of the remaining
two-point correlations on $\Sigma^{2}$ and $\bar{\Delta}_{3}$.

\begin{figure}
  \centerline{\epsfig{figure=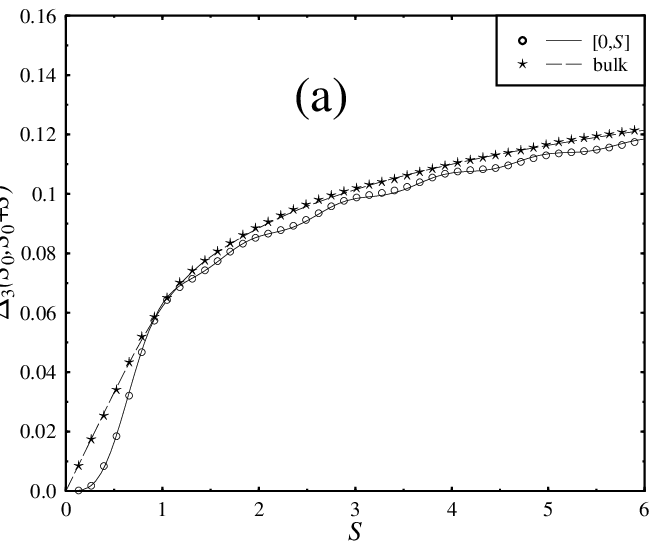,width=7.5cm}}
  \vspace*{5mm}
  \centerline{\epsfig{figure=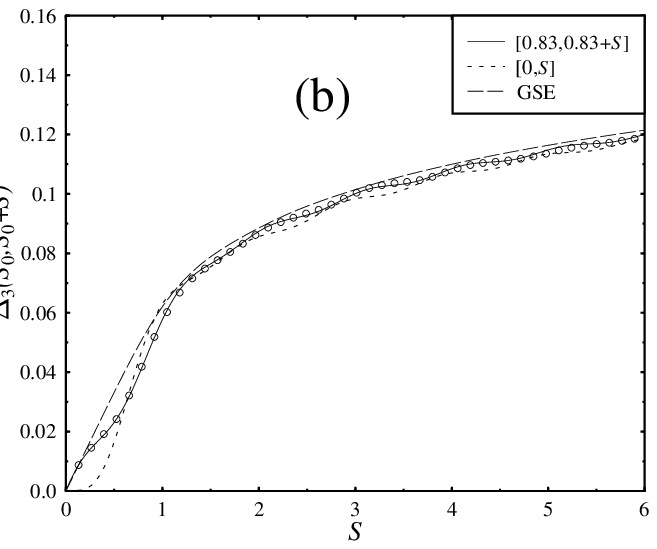,width=7.5cm}}
  \vspace*{3mm}
  \caption{The averaged spectral rigidity at the edge in intervals
    (a) $[0,S]$ and (b) $[0.83,0.83+S]$ as calculated from the lattice
    data is compared with chGSE (solid lines) and GSE predictions.}
  \label{fig5}
\end{figure}

We have also calculated the averaged spectral rigidity both at the
edge for various intervals and in the bulk. In Fig.~\ref{fig5} we plot
$\bar{\Delta}_{3}(0,S)$ and $\bar{\Delta}_{3}(0.83,0.83 +S)$ compared
with the corresponding chGSE predictions.  We see that the region
where the lattice data agree with the prediction of the chGSE is
larger than in the case of the number variance, and that the agreement
is nearly perfect.  In contrast to the case of the number variance, we
find that the averaged spectral rigidity at the edge is always smaller
than that in the bulk no matter what value of $S_{0}$ is chosen,
indicating that the spectrum at the edge is more rigid than in the
bulk.  Moreover, the difference between the edge and the bulk for the
averaged spectral rigidity is small, compared to the significant
difference in the case of the number variance.  For $S\ll 1$, the
behavior of $\bar{\Delta}_3(0,S)$ is dominated by the one-point
function and lower than the straight line $\bar{\Delta}_3=S/15$ in the
bulk case. In addition, the curve shows slight convex-concave
oscillation with the same period as the oscillations in the number
variance and the microscopic spectral density.

\section{Microscopic unfolding and two-point correlations}
\label{sec4}

As noted in the previous section, both the number variance and the
averaged spectral rigidity at the spectrum edge show oscillations with
the same period as the microscopic spectral density. We now wish to
clarify whether these oscillations are simply the manifestation of the
inhomogeneity of the microscopic spectral density, or are intrinsic in
the two-point correlations at the edge. To this end, we first need to
separate the variation of the spectral density from the fluctuations
on a smaller scale than the microscopic one. This procedure is known
as unfolding.  We therefore unfold the eigenvalues of all spectra at
the edge using the ensemble average of the staircase function,
\begin{equation}
  \label{eq4.1}
  {\bar N}(z_i)=\int_0^{z_i}dz\rho_1(z)\;.  
\end{equation}
Here, $z_{i}$ is the $i$-th eigenvalue measured in terms of the local
mean level spacing.  Since theoretical results for the staircase
function and the microscopic spectral density are available, we can
use them directly to unfold the data.  The function $\rho_{1}(z)$ is
given by Eq.~(\ref{eq2.5}).

\begin{figure}
  \centerline{\epsfig{figure=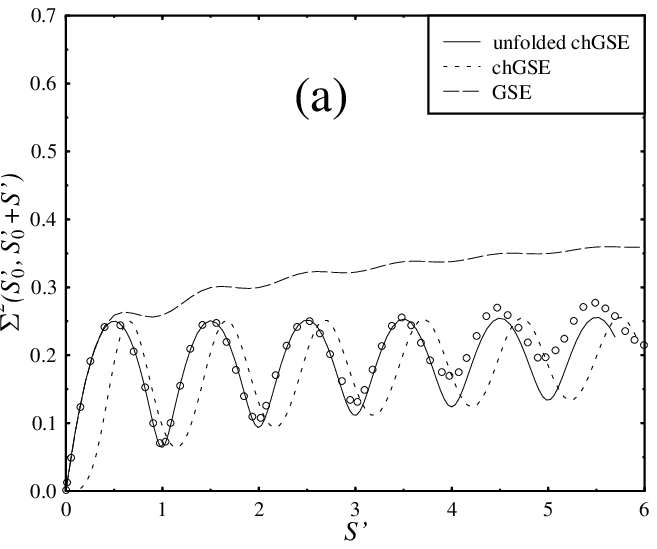,width=7.5cm}}
  \vspace*{5mm}
  \centerline{\epsfig{figure=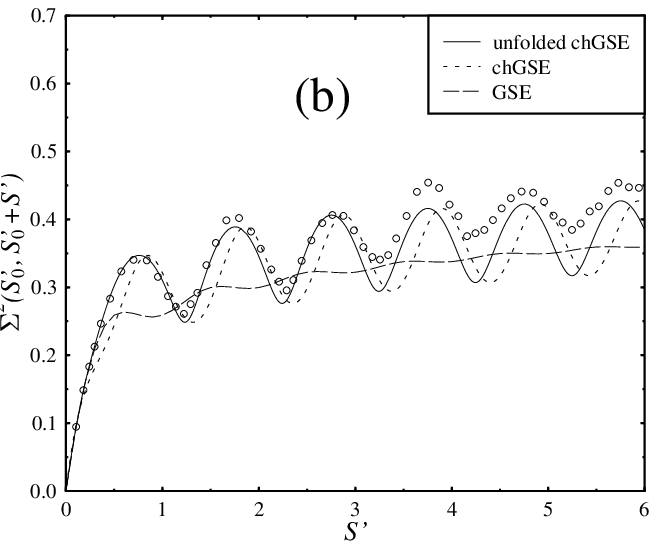,width=7.5cm}}
  \vspace*{3mm}
  \caption{The number variance calculated from the  unfolded lattice
    data at the edge in an interval (a) $[0,S']$ and in an interval
    (b) $[0.75, 0.75+S']$ corresponding to that in
    Fig.~\protect\ref{fig4}(b).}
  \label{fig6}
\end{figure}

The microscopic unfolding procedure now works as follows.  We
introduce a variable transformation from $z$ (which is on the
microscopic scale, cf.\ Eq.~(\ref{eq2.15})) to a new variable $z'$,
\begin{equation}
  \label{eq4.2}
  z\to z'=\int_{0}^{z}dt\rho_{1}(t)\equiv \omega(z)\;.
\end{equation}
We denote the inverse of $\omega(z)$ by $\Omega(z')$,
\begin{equation}
  \label{eq4.3}
  z=\Omega(z')\;.  
\end{equation}
In terms of the new variable $z'$, the microscopic spectral density
becomes 
\begin{equation}
  \label{eq4.4}
  \rho'_{1}(z')=\rho_{1}(z)\frac{dz}{dz'}=\rho_{1}\left(\Omega(z')
  \right) \cdot \frac{1}{\rho_{1}\left(\Omega(z')\right)}=1  
\end{equation}
as it should be.  The two-point correlation function becomes
\begin{eqnarray}
  \label{eq4.5}
  \rho'_2(z'_1,z'_2)&=&\rho_2(z_1,z_2)\frac{1}{dz'_1/dz_1 \cdot 
    dz'_2/dz_2}\nonumber\\
  &=&\frac{\rho_2(\Omega(z'_1),\Omega(z'_2))}{\rho_1(\Omega(z'_1))
    \rho_1(\Omega(z'_2))}\;,  
\end{eqnarray}
and the two-point cluster function is
\begin{equation}
  \label{eq4.6}
  \tau'_{2}(z'_{1},z'_{2})=1-\rho'_{2}(z'_{1},z'_{2})
  =\frac{\tau_{2}(\Omega(z'_{1}),\Omega(z'_{2}))}{\rho_{1}(\Omega(z'_{1}))
    \rho_{1}(\Omega(z'_{2}))}\;.  
\end{equation}
The number variance in an interval $[0,S']$ on the unfolded
scale is given by
\begin{eqnarray}
  \label{eq4.7}
  \Sigma'^{2}(0,S')&\equiv&S'-\int_{0}^{S'}\int_{0}^{S'}dz'_{1}dz'_{2} 
  \tau'_{2}(z'_{1},z'_{2})\nonumber\\
  &&\hspace*{-40pt}=\int_{0}^{\Omega(S')}dz\rho_{1}(z)-
  \int_{0}^{\Omega(S')}\int_{0}^{\Omega(S')}dz_{1}dz_{2}
  \tau_{2}(z_{1},z_{2})\nonumber\\
  &&\hspace*{-40pt}=\Sigma^{2}(0,\Omega(S'))\;.
\end{eqnarray}
But for the spectral rigidity, $\bar{\Delta'}_{3}(0,S')\ne
\bar{\Delta}_{3}(0,\Omega(S'))$.  Instead, 
\begin{eqnarray}
  \label{eq4.8}
  \bar{\Delta'}_{3}(0,S')&\equiv&\frac{S'}{15}+\frac{2}{S'}
  \int_0^{S'}dz'_1z'_1\int_0^{z'_1}dz'_2\tau'_{2}(z'_1,z'_2) \nonumber\\
  &&\hspace*{-45pt}-\frac{1}{S'^4}\int_0^{S'}\int_0^{S'}dz'_1dz'_2
  \tau'_2(z'_1,z'_2)[{S'}^3(z_1'+z_2')\nonumber\\
  &&\hspace*{-45pt}\phantom{+\frac{1}{S'^4}\int_0^{S'}}
  -4{S'}^2z_1'z_2'+3S'z_1'z'_2(z_1'+z_2')-3{z_1'}^2{z_2'}^2]\nonumber\\
  &&\hspace*{-55pt}=\frac{S'}{15}+\frac{2}{S'}\int_0^{\Omega(S')}dz_1
  \omega(z_1)\int_{0}^{z_1}dz_2\tau_2(z_1,z_2)\nonumber\\
  &&\hspace*{-45pt}+\frac{1}{S'^{4}}\int_{0}^{\Omega(S')}
  \int_{0}^{\Omega(S')}dz_{1}dz_{2}\tau_{2}(z_{1},z_{2})
  \{3\omega(z_1)^2\omega(z_2)^2\nonumber\\
  &&\hspace*{-10pt}-3S'\omega(z_1)\omega(z_2)[\omega(z_1)+
  \omega(z_2)]\nonumber\\
  &&\hspace*{-10pt}+4{S'}^2\omega(z_1)\omega(z_2)-
  {S'}^3[\omega(z_1)+\omega(z_2)]\}\;.
\end{eqnarray}
 
\begin{figure}
  \centerline{\epsfig{figure=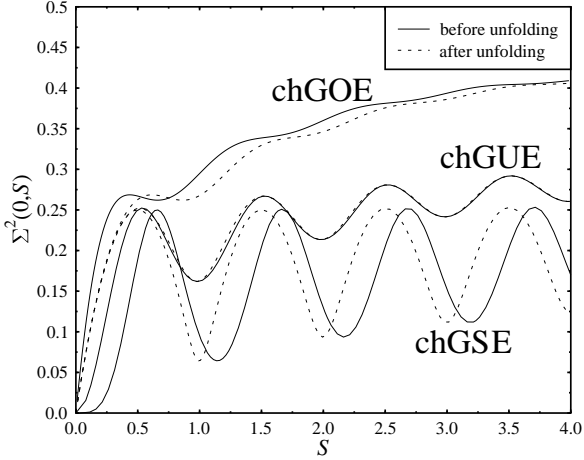,width=75mm}}
  \vspace*{3mm}
  \caption{The number variance of the chGOE, chGUE, and chGSE before
    and after unfolding in an interval $[0,S]$ at the spectrum edge.}
  \label{fig7}
\end{figure}

Figure~\ref{fig6} shows the number variance at the edge for the
unfolded data in the intervals $[0,S']$ and $[0.75,0.75+S']$
corresponding to the intervals $[0,S]$ and $[0.83,0.83+S]$ before
unfolding as in Fig.~\ref{fig4}, compared with the theoretical
prediction of Eq.~(\ref{eq4.7}) and its generalization to
$[S'_{0},S'_{0}+S']$.  On the unfolded scale, we see that the lattice
data agree with the theory in the region $0<S'+S'_{0}<4$.  As
expected, for $S'\ll 1$ the number variance now returns to the normal
case, i.e., to the straight line $\Sigma'^{2}\sim S'$.  For $S'>1$,
however, the curves still show oscillations with almost the same
amplitude as before unfolding.  Also, they still depend strongly on
the value of $S'_{0}$.  The only difference is a small shift along
$S'$.  We therefore come to the conclusion that this kind of
oscillation is intrinsic in the two-point correlations, rather than a
simple manifestation of the inhomogeneity of the microscopic spectral
density. In Fig.~\ref{fig7}, we compare the chGSE prediction with
those of the chGOE and chGUE under the same conditions, i.e.,
$N_{f}=0$ and $\nu=0$.  One can also see oscillations in the chGOE and
chGUE curves, although they are weaker than those of the chGSE.

\begin{figure}[t]
  \centerline{\epsfig{figure=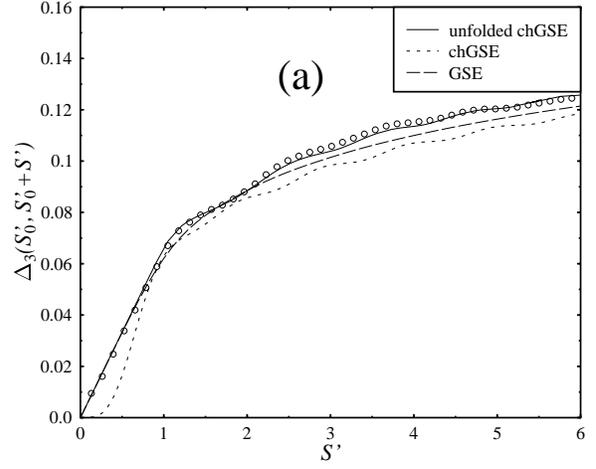,width=7.5cm}}
  \vspace*{5mm}
  \centerline{\epsfig{figure=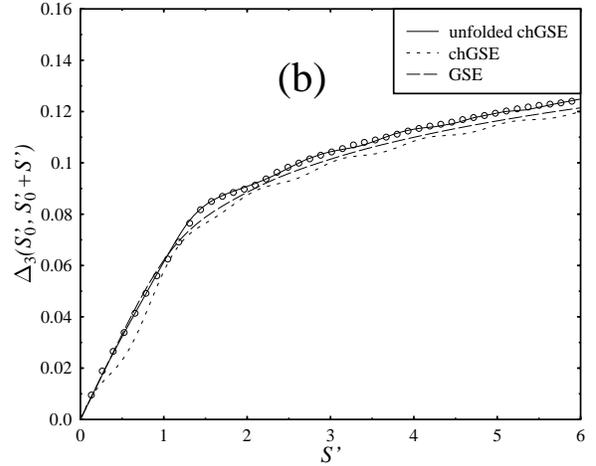,width=7.5cm}}
  \vspace*{3mm}
  \caption{The averaged spectral rigidity calculated from the unfolded
    lattice data at the edge in an interval (a) $[0,S']$ and in an
    interval (b) $[0.75, 0.75+S']$ corresponding to that in
    Fig.~\protect\ref{fig5}(b).} 
  \label{fig8}
\end{figure}

We also calculated the averaged spectral rigidity at the edge for the
unfolded data and found good agreement with the corresponding chGSE
curves, as shown in Fig.~\ref{fig8}.  We also found that after
unfolding the $\bar{\Delta}_{3}$ is larger than that before unfolding
and even larger than the GSE result for large $S'$.  This might imply
that the smaller rigidity (compared to the GSE) before unfolding is
mainly a manifestation of the oscillations in the one-point function.
The picket-fence-like distribution of $\rho_{1}$ makes the spectra
very rigid.  As before unfolding, a convex-concave oscillation is
seen.

In Ref.~\cite{Naga95}, Nagao and Forrester also define an unfolding
procedure.  Although their definition looks quite different from ours,
one can show that they are essentially identical. At first glance,
their unfolding seems to remove the global fluctuations of the
spectral density defined in terms of the original scale (cf.\ 
Eq.~(7.7) of Ref.~\cite{Naga95}):
\begin{equation}
  \label{eq4.9}
  x\to X=\int_{0}^{x}dx'R_{1}(x')\;.
\end{equation}
However, to have a finite $X$, one has to consider small $x$ so that
$x\sim 1/N$ if $N\to \infty$. This corresponds to our rescaling
$\lambda\to z=2N\Sigma\lambda/\pi$ before the unfolding of
Eq.~(\ref{eq4.1}).  We must emphasize that because of our microscopic
unfolding, the fluctuations on the scale of the mean level spacing
($z\sim 1$) have been changed.  Only on the ``sub-microscopic'' scale
($z\ll 1$) do the fluctuations remain the same as before unfolding.
For example, the level-repulsion law on the scale of the mean-level
spacing must be the same before and after unfolding. But the
``long-range'' statistics, like the number variance
$\Sigma'^{2}(S'_{0},S'_{0}+S')$, is different from
$\Sigma^{2}(S_{0},S_{0}+S)$, as seen in Fig.~\ref{fig6}.
Nevertheless, our results show that for an ensemble of spectra with a
homogeneous spectral density, the number variance can show strong
oscillations.  Therefore, this kind of oscillation in the number
variance can not be attributed to the oscillations in the microscopic
spectral density but is inherent in the two-point correlations.

\section{On the spacing distribution}
\label{sec5}

We now turn to the short-range statistics. As usual, we define
\begin{equation}
  \label{eq5.1}
  E(L_{0},L_{0}+L)=
  \mathop{\int \cdots \int}_{{\rm out}} d\lambda_{1}\cdots d\lambda_{N}
  \rho(\lambda_{1},\ldots,\lambda_{N})\:,  
\end{equation}
where ``out'' stands for $[0,L_{0}]$ and $[L_{0}+L,\infty)$. Known as
the ``hole'' probability, $E(L_{0},L_{0}+L)$ is the probability that
the interval $[L_{0},L_{0}+L]$ is free of eigenvalues.  A related
probability density, $F(L_{0},L_{0}+L)$, is defined as
\begin{eqnarray}
  \label{eq5.2}
  F(L_{0},L_{0}+L)=&&\nonumber\\
  &&\hspace*{-70pt}N\mathop{\int \cdots \int}_{{\rm out}} d\lambda_{2}
  \cdots d\lambda_{N}\rho(L_{0}+L,\lambda_{2},\ldots,\lambda_{N})\;.  
\end{eqnarray}
\begin{figure}
  \centerline{\epsfig{figure=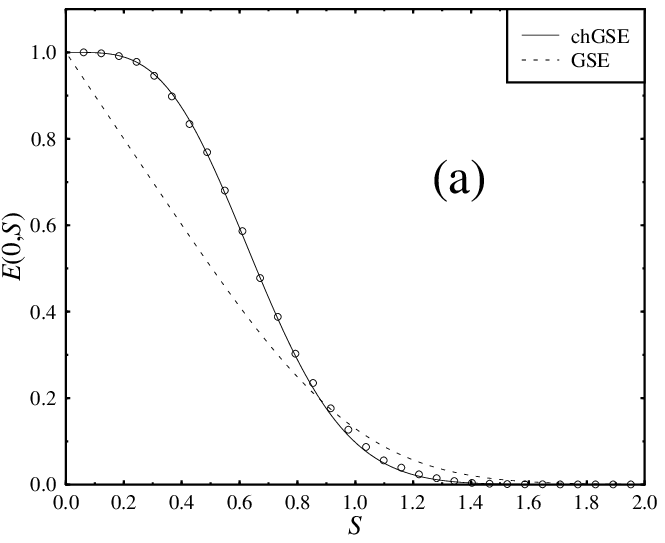,width=7.5cm}}
  \vspace*{5mm}
  \centerline{\epsfig{figure=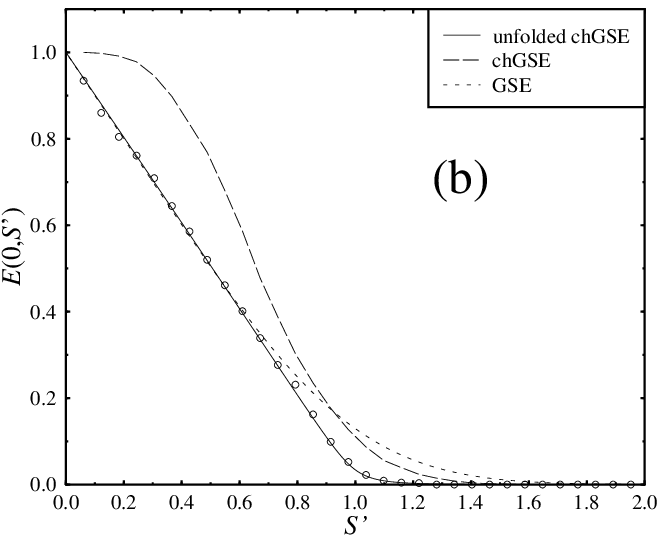,width=7.5cm}}
  \vspace*{3mm}
  \caption{The hole probability $E(0,S)$ (a) before and (b) after
    microscopic unfolding, compared with the corresponding chGSE
    predictions.  The circles are the lattice data.}
  \label{fig9}
\end{figure}

\noindent $F(L_{0},L_{0}+L)dL$ is the probability that the interval
$[L_{0},L_{0}$ $+L]$ is free of eigenvalues and that an eigenvalue is
found in $[L_{0}+L,L_{0}+L+dL]$.  The nearest-neighbor spacing
distribution function is defined as
\begin{eqnarray}
  \label{eq5.3}
  P(L_{0},L_{0}+L)=&&\nonumber\\
  &&\hspace*{-75pt}N(N-1)\mathop{\int \cdots \int}_{{\rm out}}
  d\lambda_{3}\cdots d\lambda_{N}\rho(L_{0},L_{0}+L,\lambda_{3},
  \ldots,\lambda_{N})\nonumber\\
  &&
\end{eqnarray}
so that $P(L_{0},L_{0}+L)dL_{0}dL$ is the probability that two
eigenvalues are located in $[L_{0},L_{0}+dL_{0}]$ and
$[L_{0}+L,L_{0}+L+dL]$, respectively, and that there is no other
eigenvalue between them.  Again, we rescale $L\to S=2N\Sigma L/\pi$ to
move to the spectrum edge.  We then have
\begin{eqnarray}
  \label{eq5.4}
  E(S_0,S_0\!+\!S)&=&\lim_{N\to\infty}E\left(\frac{\pi S_0}{2N\Sigma},
    \frac{\pi S_0}{2N\Sigma}\!+\!\frac{\pi S}{2N\Sigma}\right)\;,
  \nonumber\\
  && \\
  \label{eq5.5}
  F(S_{0},S_{0}\!+\!S)&=&\lim_{N\to \infty}\frac{\pi}{2N\Sigma}
  F\left(\frac{\pi S_{0}}{2N\Sigma},\frac{\pi S_{0}}{2N\Sigma}
    \!+\!\frac{\pi S}{2N\Sigma}\right)\;,\nonumber\\
  && \\
  \label{eq5.6}
  P(S_{0},S_{0}\!+\!S)&=&\!\!\lim_{N\to\infty}\frac{\pi^2}{(2N\Sigma)^2} 
  P\left(\frac{\pi S_{0}}{2N\Sigma},\frac{\pi S_{0}}{2N\Sigma}
    \!+\!\frac{\pi S}{2N\Sigma}\right)\!.\nonumber\\
  &&
\end{eqnarray}
For the special case $S_{0}=0$ we obtain, using the results of
Ref.~\cite{Forr93}, 
\begin{eqnarray}
  \label{eq5.7}
  E(0,S)&=&\frac{\pi}{\sqrt{2}}S^{1/2}\exp(-\pi^2S^2/2)
  I_{-1/2}(\pi S)\;,\\ 
  \label{eq5.8}
  F(0,S)&=&\frac{\pi^3}{\sqrt{2}}S^{3/2}\exp(-\pi^2S^2/2)I_{3/2}(\pi S)\;,
\end{eqnarray}
where $I$ denotes the modified Bessel function.  (The two quantities
are related by $F(0,S)=-dE(0,S)/dS$.)  However,
\begin{equation}
  \label{eq5.9}
  P(0,S)\equiv 0
\end{equation}
due to the repulsion between an eigenvalue and the origin inherent in
the joint distribution function, Eq.~(\ref{eq2.1}).  By definition,
$F(0,S)$ is the probability density of the smallest eigenvalue.

\begin{figure}
  \centerline{\epsfig{figure=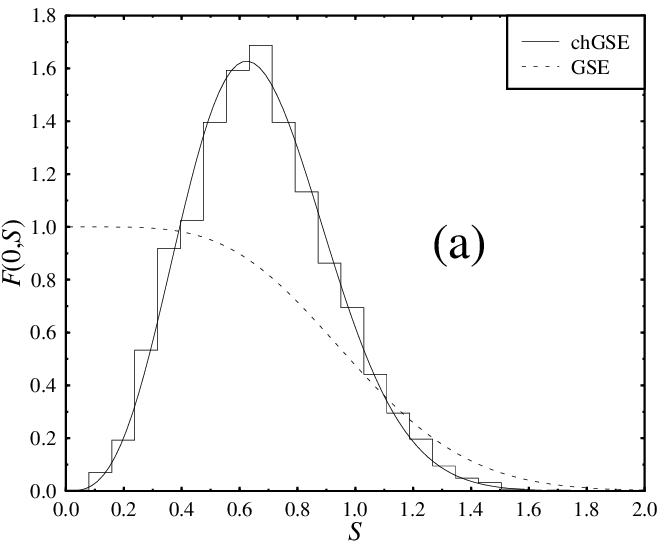,width=7.5cm}}
  \vspace*{5mm}
  \centerline{\epsfig{figure=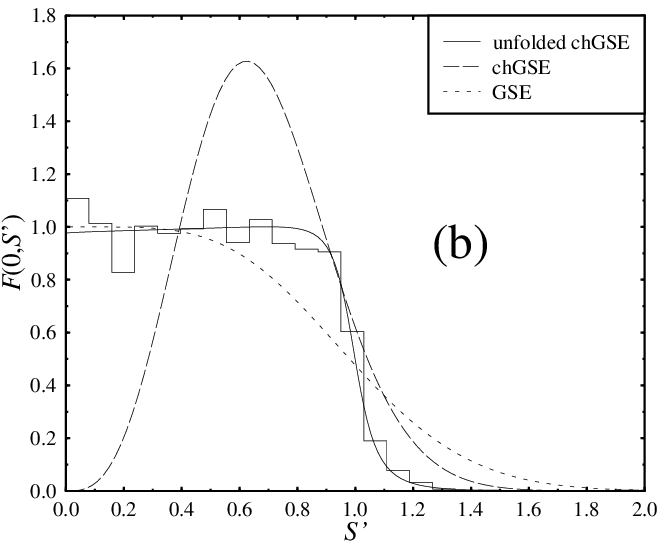,width=7.5cm}}
  \vspace*{3mm}
  \caption{The distribution of the smallest eigenvalue $F(0,S)$ (a)
    before and (b) after microscopic unfolding, compared with the
    corresponding chGSE predictions.  The histograms represent the
    lattice data.}
  \label{fig10}
\end{figure}

Figure~\ref{fig9} shows the results for $E(0,S)$ calculated from the
lattice data before and after microscopic unfolding. We again
observe nice agreement with the corresponding chGSE predictions.
Similar results for $F(0,S)$ are shown in Fig.~\ref{fig10}. (Note that
$F(0,S)$ plotted in Fig.~\ref{fig10}(a) corresponds to the
distribution of the smallest eigenvalue obtained in
Ref.~\cite{Berb97a}.) {}From Fig.~\ref{fig9} one can see that, after
microscopic unfolding, the shape of $E(0,S)$ becomes closer to that of
the usual GSE.  The eigenvalues of the Dirac operator always occur in
pairs $\pm \lambda_{k}$.  Therefore, the function $F(0,S)$ may also be
viewed as the nearest-neighbor spacing distribution at the origin if
one defines $2S$ as the spacing between the pair $\pm\lambda_{\rm
  min}$ in units of the mean level spacing.  Since the spacing
distribution depends on correlators of all order, the nice agreement
with the chGSE predictions seen in these two figures provides evidence
for the universality of higher-order correlations.  For the generic
$P(S_{0},S_{0}+S)$ defined in Eq.~(\ref{eq5.6}) with $S_{0}\ne 0$
there is, to the best of our knowledge, no theoretical expression
available.  Of course, it can be constructed from the lattice data.
We found that for a generic value of $S_{0}$, the histogram (which is
not shown here) is quite different from the usual GSE prediction. It
would be interesting to investigate this problem further.

\section{Summary}
\label{sec6}

We have studied the spectral statistics of the Dirac spectrum in an
SU(2) gauge theory with staggered fermions, restricting ourselves to
an $8^{4}$ lattice, with emphasis on the long-range statistics, i.e.,
the number variance and the spectral rigidity, at the spectrum edge
near zero virtuality.  Our analysis shows that while the spectra at
the edge are more rigid than in the bulk, the fluctuations are
suppressed, on average, to the same extent as in the bulk.  The strong
oscillation in $\Sigma^{2}$ is an edge effect due to the two-point
correlations, in the sense that it cannot be removed by microscopic
unfolding. On the other hand, the larger rigidity is due to the
picket-fence-like behavior of the one-point function, and can be
reduced by microscopic unfolding.  The excellent agreement between the
lattice data and the predictions from chiral random-matrix theory
provide direct evidence for the universality of the one- and two-point
spectral correlations at the edge.  Our study of the spacing
distribution demonstrates this universality also for higher-order
correlations.

Since our study was only done for one particular lattice size and one
particular value of $\beta$, it is in order to discuss the effect of
these two parameters on our results.  The random-matrix results to
which we compare, in particular Eqs.~(\ref{eq2.5}) and (\ref{eq2.9}),
were derived in the thermodynamic limit.  Thus, the agreement between
lattice data and chRMT predictions will improve with increasing
physical volume $V$, i.e., with increasing lattice size and decreasing
$\beta$.  We have confirmed this expectation investigating lattice
data by Berbenni-Bitsch and Meyer obtained for a number of
$\beta$-values in the region $1.8\cdots 2.5$ and lattice sizes between
$6^4$ and $16^4$.  It should be emphasized that for any given value of
$\beta$ one will eventually find agreement between lattice data and
chRMT, provided that the lattice is big enough.  Of course, there are
practical constraints.  Furthermore, chRMT cannot predict a priori how
good the agreement will be for a given set of lattice parameters.
However, the criterion after Eq.~(\ref{eq3.1}) gives an estimate for
the quality of the agreement.  The mean level spacing at the spectrum
edge has to be much smaller than $1/a$, where $a$ is the lattice
constant.

After all these demonstrations of universality, one interesting
question is the one of practical applications.  One possibility is the
improvement of extrapolations to the thermodynamic limit, as shown in
Ref.~\cite{Berb97b}.  Another issue is the extrapolation to the chiral
limit in the presence of dynamical fermions.  Lattice data for such an
investigation are just becoming available, and analytical work on the
appropriate random-matrix model is in progress.

\begin{acknowledgement}
  We are grateful to M.E.\ Berbenni-Bitsch and S.\ Meyer for providing
  us with their lattice data and to A.D.\ Jackson,
  A.\ M\"uller-Groeling, A.\ Sch\"afer, J.J.M.\ Verbaarschot,
  H.A.\ Weidenm\"uller, and T.\ Wilke for stimulating discussions.
  T.W.\ acknowledges the hospitality of the MPI Heidelberg.  This work
  was supported in part by DFG grant We 655/11-2.
\end{acknowledgement}


\begin{thebibliography}{99}
\bibitem{Bohi84} O.\ Bohigas and M.J.\ Giannoni, Lec.\ Not.\ Phys.\
  {\bf 209} (Springer, Heidelberg, 1984).
\bibitem{Shur93} E.V.\ Shuryak and J.J.M.\ Verbaarschot, Nucl.\ Phys.\
  A {\bf 560}, 306 (1993).
\bibitem{Verb94a} J.J.M.\ Verbaarschot, Phys.\ Rev.\ Lett.\ {\bf 72},
  2531 (1994). 
\bibitem{Verb97} J.J.M.\ Verbaarschot, hep-th/9710114.
\bibitem{Hala95} M.A.\ Halasz and J.J.M.\ Verbaarschot, Phys.\ Rev.\
  Lett.\ {\bf 74}, 3920 (1995); 
  M.A.\ Halasz, T.\ Kalkreuter, and J.J.M.\ Verbaarschot, Nucl.\
  Phys.\ B (Proc.\ Supp.) {\bf 53}, 266 (1997).
\bibitem{Wilk97} T. Wilke, private communication.
\bibitem{Naga91} D.\ Fox and P.B.\ Kahn, Phys.\ Rev.\ {\bf 134}, B1151
  (1964); T.\ Nagao and M.\ Wadati, J.\ Phys.\ Soc.\ Jpn.\
  {\bf 60}, 3298 (1991); {\bf 61}, 78, 1910 (1992).
\bibitem{Leut92} H.\ Leutwyler and A.V.\ Smilga, Phys.\ Rev.\ D {\bf
    46}, 5607 (1992).
\bibitem{Bank80} T.\ Banks and A.\ Casher, Nucl.\ Phys.\ B {\bf 169},
  103 (1980).
\bibitem{Verb93} J.J.M.\ Verbaarschot and I.\ Zahed, Phys.\ Rev.\
  Lett.\ {\bf 70}, 3852 (1993).
\bibitem{Verb94c} J.J.M.\ Verbaarschot, Nucl.\ Phys.\ B {\bf 426}, 559
  (1994). 
\bibitem{Naga95} T.\ Nagao and P.J.\ Forrester, Nucl.\ Phys.\ B {\bf
    435}, 401 (1995).
\bibitem{Verb94b} J.J.M.\ Verbaarschot, Nucl.\ Phys.\ B {\bf 427}, 534
  (1994).
\bibitem{Chan95} S.\ Chandrasekharan and N.\ Christ, Nucl.\ Phys.\ B
  (Proc. Suppl.) {\bf 47}, 527 (1996).
\bibitem{Verb96a} J.J.M.\ Verbaarschot, Phys.\ Lett.\ B {\bf 368}, 137
  (1996).
\bibitem{Brez96} E.\ Br\'ezin, S.\ Hikami, and A.\ Zee, Nucl.\ Phys.\
  B {\bf 464}, 411 (1996).
\bibitem{Nish96}
 S.\ Nishigaki, Phys.\ Lett.\ B {\bf 387}, 139 (1996); G.\ Akemann,
 P.H.\ Damgaard, U.\ Magnea, and S.\ Nishigaki, Nucl.\ Phys.\ B {\bf
   487}, 721 (1997). 
\bibitem{Slev93} K.\ Slevin and T.\ Nagao, Phys.\ Rev.\ Lett.\ {\bf
    70}, 635 (1993). 
\bibitem{Jack96b} A.D.\ Jackson, M.K.\ \c Sener, and
  J.J.M.\ Verbaarschot, Nucl.\ Phys.\ B {\bf 479}, 707 (1996),
  Nucl.\ Phys.\ B {\bf 506}, 612 (1997); T.\ Guhr and T.\ Wettig,
  Nucl.\ Phys.\ B {\bf 506}, 589 (1997).
\bibitem{Berb97a} M.E.\ Berbenni-Bitsch, S.\ Meyer, A. Sch\"afer,
  J.J.M.\ Verbaarschot, and T.\ Wettig, hep-lat/9704018, to appear in
  Phys.\ Rev.\ Lett.
\bibitem{Jack96a} A.D.\ Jackson and J.J.M.\ Verbaarschot,
  Phys.\ Rev.\ D {\bf 53}, 7223 (1996).
\bibitem{Wett96} T.\ Wettig, A.\ Sch\"afer, and H.A.\ Weidenm\"uller,
  Phys.\ Lett.\ B {\bf 367}, 28 (1996).
\bibitem{Step96a} M.A.\ Stephanov, Phys.\ Lett.\ B {\bf 375}, 249 (1996).
\bibitem{Step96b} M.A.\ Stephanov, Phys.\ Rev.\ Lett.\ {\bf 76}, 4472
  (1996). 
\bibitem{Nowa96} M.A.\ Nowak, G.\ Papp, and I.\ Zahed, Phys.\ Lett.\ B
  {\bf 389}, 137, 341 (1996); for a review, see R.A.\ Janik, M.A.\ 
  Nowak, G.\ Papp, and I.\ Zahed, hep-th/9710103.
\bibitem{Jani97a} R.A.\ Janik, M.A.\ Nowak, and I.\ Zahed, Phys.\
  Lett.\ B {\bf 392}, 155 (1997).
\bibitem{Hala97a} M.A.\ Halasz, A.D.\ Jackson, and J.J.M.\ Verbaarschot,
  Phys.\ Lett.\ B {\bf 395}, 293 (1997); Phys.\ Rev.\ D {\bf 56}, 5140
  (1997). 
\bibitem{Hala97b} M.A.\ Halasz, J.C.\ Osborn, and J.J.M.\ Verbaarschot,
  Phys.\ Rev.\ D {\bf 56}, 7059 (1997).
\bibitem{Jani97b} R.A.\ Janik, M.A.\ Nowak, G.\ Papp, and I.\ Zahed,
  Nucl.\ Phys.\ B {\bf 498}, 313 (1997).
\bibitem{Berb97b} M.E.\ Berbenni-Bitsch, A.D.\ Jackson, S.\ Meyer,
  A. Sch\"afer, J.J.M.\ Verbaarschot, and T.\ Wettig, hep-lat/9709102.
\bibitem{Meht91} M.L.\ Mehta, {\it Random Matrices}, 2nd ed. (Academic
  Press, San Diego, 1991).
\bibitem{Meht63} M.L.\ Mehta and F.J.\ Dyson, J.\ Math.\ Phys.\ {\bf
  4}, 713 (1963).
\bibitem{Forr93} P.J.\ Forrester, Nucl.\ Phys.\ B {\bf 402}, 709 (1993).
\end{thebibliography}
\end{document}